# Bose Einstein condensation and superfluidity in an open system: theory

G. Malpuech, D. Solnyshkov

*LASMEA, CNRS and University Blaise Pascal, 63177 Aubière cedex France.*

We propose a new theoretical formalism which describes the Bose Einstein condensation of weakly interacting bosons with finite life time interacting with a thermal bath. We show that if a quasi-thermal distribution function of particles is achieved, the elementary excitations of the condensate show a linear spectrum characteristic of a superfluid, with a renormalized sound velocity with respect to the equilibrium case.

The Bose Einstein Condensation (BEC) of cavity exciton-polaritons [1] (polaritons) became in the last years a very intense and exciting field of research. After the first clear evidences of the achievement of polariton BEC [2,3], an intense race started toward the observation of the fascinating related phenomena. One of this effects is superfluidity [4]. From the Landau criterion [5], a fluid moves without being affected by mechanical friction if the dispersion of its elementary excitations depends linearly or sub-linearly on its velocity or wave vector. It was then shown by Bogoliubov [6] that a BEC of weakly interacting bosons should indeed present such a linear dispersion $\hbar c_s k$, where $c_s = \sqrt{\mu/m}$ is the speed of sound in the media, $\mu$ being the chemical potential and $m$ the mass of particles. Such a linear dispersion has not been observed in the experiments [2,3] and several explanations have been proposed. One is based on the impact of the structural disorder which at a moderate condensate density should be able to destroy the superfluid behaviour [7] of the polariton gas. Another explanation relies on the intrinsic nature of a polariton condensate which, being made of particles having a finite life time, requires to be constantly pumped. Two different groups using two different approaches have concluded that Bose condensation in an open system should not be accompanied by the formation of a Bogoliubov-like spectrum of excitations [8,9]. The recent claiming [10] of observation of such a linear spectrum in a GaAs-based cavity, less affected by disorder than the CdTe structure of [2,3], seems to argue in favour of the first interpretation. On the other hand, the debate is not closed since, from our point of view, there is no satisfactory model developed so far which is able to describe BEC in an open system. Such description clearly represents a strong theoretical challenge. Two main approaches have been mainly used so far. The first type is based on the description of the system in terms of one time density matrix. The most simple and popular example of this approach is the semi-classical Boltzman equations used by many authors to describe polariton condensation [11,12,13,14]. The Boltzmann equations describe the evolution of the mean value of the diagonal element of the polariton density matrix which in other words is the occupation of the states characterised by a given wave vector $n(\vec{k}, t)$. These equations can be rigorously derived from the Hamiltonian describing weakly interacting bosons using the so-called Born Markov approximation together with the assumption of a homogeneous distribution of particles in real space. They allow to take into account properly the life time, non-resonant pumping, spontaneous and stimulated scattering, and interaction with a thermal bath of phonons. A more evolved version of the Boltzmann equation is the so-called Ueling-Uhlenbeck equation, whose rigorous derivation can be found in [15]. This equation can be seen as space-dependent Boltzmann equations $n(\vec{k},\vec{r},t)$ which is made possible by the assumption of a finite coherence length and a wavelet description of the polariton state. Master equation approaches have also been developed [16], which allow to describe the type of statistics (thermal or coherent) which develops in the condensate and to calculate the second order coherence dynamics and decoherence time of the condensate [17]. This approach has been recently improved using a Wigner function description [18]. The common feature of all these models is that they do not allow to fully describe the coherent evolution of a condensate wave function and the renormalisation of the dispersion relation induced by the strong occupation of a single quantum state. In other words, they do not allow to conclude about the superfluid behaviour of a condensate. A completely opposite description is based on the Gross Pitaevskii (GP) [5] equation which is a non-linear Schrödinger equation describing the motion of a condensate wave function $\psi(\vec{r},t)$. As such, it does not include dissipation processes like the coupling with phonons, or spontaneous scattering. It allows however to describe spatial propagation effects and renormalisation of the polariton dispersion due to the condensation effect and the normal to superfluid transition for an equilibrium condensate. It has therefore been used to describe properties of polariton condensate once this condensate is formed [7, 19, 20]. Wouters and Carusotto proposed a model [9], where the generalized GP equation takes into account the radiative decay of the wave function and the coupling to a reservoir which receives an external pumping of particles. This model is very similar to Ginzburg Landau equations recently used by J. Keeling [21] and both have been extensively used in the past two years [22,23].

From our point of view, these interesting approaches suffer of several weaknesses. One is the absence of a Langevin fluctuating term in the GP equation, which does not allow to properly describe spontaneous scattering towards the ground state. The second one is the complete absence of any thermalization process, which implies that scattering towards the ground state is similar to the one towards excited states [22]. In other words, a system with an infinite temperature is described, which, in the framework of the description of BEC phenomena, does not look suitable.

In this paper, we propose a new model which intends to make a bridge between the Boltzmann and GP description of the condensation process. It allows to correct the deficiencies of the previous models, taking into account most of the relevant physical phenomena, such as the temperature (thermalisation), finite life time of particles, real space dependence, and fluctuations. As a second step, we consider the behaviour of this model assuming the formation of a homogenous normal gas of polaritons. We assume this gas to be characterised by an effective temperature $T_X$. We deduce analytically two critical conditions for the formation of a steady condensate in the ground state. If these conditions are fulfilled, we show that the elementary excitations of the condensate are Bogoliubov-like and that the condensed part of the system should behave as superfluid, despite the finite life time and pumping.

We consider that polaritons in a normal uncondensed state are characterised by a density $n(\vec{k},\vec{r},t)$, whose dynamics is described by the Uehling-Uhlenbeck equation derived by Gardiner and Zoller [15]. This means that for normal polaritons the coherence length is finite, limited to the size of a spatial cell. On the opposite, the condensate part is described by a coherent wave function

$$\psi(\vec{r},t) = \frac{1}{(2\pi)^2}\int \psi(\vec{k},t)\exp(i\vec{k}\vec{r})d\vec{k} \qquad (1)$$

which implies an infinite coherence length.

The dynamics of the condensate is described by the GP equation with the coupling to the normal states and the coupling to the phonon bath. It reads:

$$i\dot{\psi}(\vec{k},t) = -i\frac{\gamma_k}{2}\psi(\vec{k},t) - \frac{E(k)}{\hbar}\psi(\vec{k},t) + \alpha\int\left(|\psi(\vec{r},t)|^2 + \sum_{k'}n(\vec{k}',\vec{r},t)\right)\psi(\vec{r},t)\exp(i\vec{k}\vec{r})d\vec{r} + \\ +i\frac{1}{2}\left(W_{in}^k - W_{out}^k\right)\psi(\vec{k},t) + i\frac{1}{2}W_{in}^k f(t) \qquad (2)$$

where $\gamma_k$ is the decay rate of a polariton state $k$, $E(k)$ is the polariton dispersion. $\alpha$ is the matrix element of the polariton-polariton interaction, $f(t)$ is a fluctuating Langevin force responsible for spontaneous scattering which we do not detail here.

$$W_{in}^k = \sum_{k',r} W^1_{k'\to k,r} n(\vec{k}',\vec{r},t) + \sum_{k'} W^2_{k'\to k,r}\left|\psi(\vec{k}',t)\right|^2 \qquad (3)$$

$$W_{out}^k = \sum_{k',r} W^3_{k\to k',r}\left(1 + n(\vec{k}',\vec{r},t)\right) + \sum_{k'} W^4_{k\to k',r}\left(\left|\psi(\vec{k}',t)\right|^2 + f(t)\right) \qquad (4)$$

$W^i_{\vec{k}\to\vec{k}',r}$ are the semi-classical Boltzman rates. The subscript $i$ means that they are all these scattering rates are formally different, but we are not going to give a detailed expression for all of them. They can include phonon-assisted and polariton-polariton scattering. Phonon-assisted scattering rates are similar to the standard rates calculated for homogeneous systems and do not depend on $r$, which is not the case considering polariton-polariton interaction. An essential contribution to the condensate thermalisation is due to scattering processes involving an interaction between a condensed particle and a normal particle. Indeed, the excitons, like normal particles, are often well thermalised close to the lattice temperature. Their interaction with the condensed particles provides a very efficient thermalisation mechanism for the latter ones. This mechanism can be strongly reduced in case when the normal and condensed particles show different spatial localisation because the polariton-polariton interaction involves a contact between the excitonic components of the polaritons. This situation is likely to take place in a realistic experiment with a finite spot size, because the condensed particles are normally much lighter than the exciton-like normal states. They can therefore propagate much more rapidly and leave the region where the excitonic reservoir is constantly refilled by the external pumping. In order to properly take into account such effect and preserve the locality of the polariton-polariton interaction, the scattering rates corresponding to this process should be proportional to the number of condensed particles present in a spatial cell with a given wave vector, known with an uncertainty inversely proportional to the size $\Delta R$ of the cell, which reads $n_c(k,r) = \left| \int_{cell(r)} \psi(r') \exp(ikr') dr' \right|^2$. The spatial integration takes place over the spatial cell centred on $r$, whose size is given by the coherence length of the condensate. In that framework the part of the polariton-polariton scattering rate associated with this process reads:

$$W^{pol}_{\vec{k} \to \vec{k}',r} = M_0 \sum_{\vec{k}''} n_c(\vec{k}'',\vec{r},t)\left(2 + n(k+k''-k',r) + |\psi(k+k''-k',t)|^2\right)\delta(E(k'')+E(k)-E(k')-E(k+k''-k'))$$
(5)

where $M_0$ is the matrix element of the polariton-polariton interaction, which formally depends on wave vector through the excitonic fraction of the various states and $\delta$ is the Dirac function which ensures energy conservation during scattering processes. We assume that these energies are the ones of the unperturbed dispersion $E(k)$. This means that polariton-polariton interaction induces a rigid shift of the dispersion for both normal and condensed particles which is physically consistent. On the other hand we make an approximation by not taking into account the possible change of the shape of the condensate dispersion while calculating the scattering rates. The equation for the reservoir reads :

$$\dot{n}(\vec{k},\vec{r},t) = P(\vec{k},\vec{r},t) - \frac{n(\vec{k},\vec{r},t)}{\tau_k} + \frac{dE(k)}{dk}\frac{\vec{k}}{k}\nabla_r n(\vec{k},\vec{r},t) +$$
$$\sum_{k'} W^5_{\vec{k}' \to k,r} n(\vec{k}',\vec{r},t)(1+n(\vec{k},\vec{r},t)) - \sum_{k'} W^6_{k \to k',r} n(\vec{k},\vec{r},t)(1+n(\vec{k}',\vec{r},t)) \qquad (6)$$
$$- \sum_{k'} W^7_{k \to k',r} n(\vec{k},\vec{r},t)\left(|\psi(\vec{k}',t)|^2 + f(t)\right) + \sum_{k'} W^8_{k' \to k,r} |\psi(\vec{k}',t)|^2 (n(\vec{k},\vec{r},t)+1)$$

Equations (1-6) represent a closed set of equations which can be solved numerically to describe any relevant physical situations. The stimulated scattering towards a normal state is limited by the finite coherence length. If this coherence length is taken small enough, the stimulated scattering terms for the normal part become negligible compared with the spontaneous scattering ones. As a result, the equilibrium distribution of uncondensed particles follows a Boltzmann distribution rather than a Bose one. The stimulated scattering towards the condensed states has no real space dependence because of the large coherence length we assume (equal to the total system size). This is in sharp contrast with the model [9], where the stimulated scattering is written in direct space, has no wave vector dependence, and cannot include any thermalisation processes, for instance.

In what follows, we are not going to solve the system (1-6) numerically, which will be the topic of future extended works. We will rather consider a limit case where analytical derivations can be performed. We consider the limit of an infinitely large homogeneous system. One of the consequences is that the spontaneous scattering term vanishes from the equation (2). We look for a steady state solution of our system of equations. We moreover assume the achievement of a homogeneous, Boltzmann-like distribution function for the particles in the normal state.

$$n(\vec{k},\vec{r}) = n_0 \exp(-E(k)/k_b T_X) \qquad (7)$$

$n_0$ is related to the density of particles present in the normal state. It is therefore related to the external pumping intensity. $T_X$ is the polariton temperature. This type of assumption is based on i) several experimental works reporting the achievement of a quasi-thermal distribution function for the polaritons [3,24,25,26,27]; ii) numerous simulations based on semi-classical Boltzmann equations which for several years had predicted that this type of situation should occur and have then been confirmed the above mentioned measurements. The polariton temperature exceeds the temperature of the phonon bath because of the finite life-time of particles. On the other hand, we want to insist that the existence of such a temperature, which can be in some cases very close to the lattice temperature, makes relevant the use of thermodynamics to describe some features of polariton condensates. In such a regime, the critical density of condensation and its dependence on the system parameters can be evaluated within the thermodynamic approach with a reasonably good accuracy [25].

For the condensate wave function we look for a solution of the type:

$$\psi(\vec{r},t) = \psi_0 \exp(-i\mu t) \qquad (8)$$

We are now going to calculate the scattering rates. For simplicity we consider the phonon-assisted scattering only. This does not mean the polariton-polariton interaction is neglected. This type of interaction plays an essential role in the achievement of a thermal distribution function for the polariton gas and is therefore implicitly accounted for in the equation (7). In this framework the scattering rates present in the equations (3) and (4) read:

$$W_{in}^k = \frac{L^2}{4\pi^2} \int W_0 n_0 \exp(-E(k')/k_b T_X)(1 + n_{ph}(E(k') - E(k))) d\vec{k}' \qquad (9)$$

$$W_{out}^k = \frac{L^2}{4\pi^2} \int W_0 \left(1 + n_0 \exp(-E(k')/k_b T_X) n_{ph}(E_X(k') - E(k))\right) d\vec{k}' \qquad (10)$$

where $L^2$ is a normalisation area which disappears when the integration is performed. $W_0$ characterises the strength of the polariton-phonon interactions. It actually depends on $\vec{k}'$ and

$\vec{k}$. However, we are going to neglect this dependence at this stage. $n_0$ is proportional to the number of particles in the normal state. In order to simplify the algebra we also consider a Boltzmann distribution of phonons, which allows to write :

$$\Delta W^k = W_{in}^k - W_{out}^k = W_0 \left( n_0 g_1(T_X) - g_1(T) \exp(E(k)/k_b T) \right) \tag{11}$$

$$\text{where } g_1(T) = \frac{L^2}{4\pi^2} \int \exp(-E(k')/k_b T) d\vec{k}' \tag{12}$$

which cannot be computed analytically because of the non-parabolic polariton dispersion. If $T = T_X$, (11) is negative except for the ground state where it cancels. The condensation is therefore possible only for infinitely long life time. Therefore, the condensation for decaying particles is possible only in case of an excess temperature of the polariton gas with respect to the one of the phonon gas, as noticed by Imamoglu in his seminal paper [28]. One should also notice that $\Delta W^k$ is a decaying function of $k$ which favours the condensate formation in $k = 0$, contrary to other models [9], where the thermalisation is absent. Another extremely important quantity is the decay rate which in a microcavity depends on the wave vector. The decay rate is typically governed by the photonic fraction and the cavity photon life time. The decay therefore decreases versus $k$, which is at the origin of the so-called bottleneck effect [11]. The life time is longer for excited states than for the ground states. In some cases this leads to an accumulation of particles in the excited states rather than in the ground state. At zero detuning between the photon and exciton mode, and close to $k = 0$, the balance between the scattering rate and decay time reads :

$$\Delta W^k - \gamma_k = W_0 \left( n_0 g_1(T_X) - g_1(T) \right) - \gamma_0 - \frac{\hbar^2 k^2}{2m} \left( \frac{W_0 g_1(T)}{k_b T} - \frac{\gamma_0}{\hbar \Omega} \right) \tag{13}$$

From the formula (1), one can see that several regimes can be achieved depending on the sign of the quantity

$$\beta = \frac{W_0 g_1(T)}{k_B T} - \frac{\gamma_0}{\hbar \Omega} \tag{14}$$

If $\beta > 0$, the system is in the classical situation where the gain decays versus wave vector. The ground state is favoured with respect to the higher energy states. This situation is achieved, if the effective scattering rate $W_0$ and density $n_0$ are large enough compared with the decay rate $\gamma_0$. A large temperature also favours the achievement of such regime, since $g_1(T)/T$ is a growing function of $T$. Moreover, the increase of the Rabi splitting favours the achievement of $\beta > 0$ as well. In general, a steady state stationary solution will form in the ground state if the critical condition

$$W_0 \left( n_0 g_1(T_X) - g_1(T) \right) - \gamma_0 = 0 \tag{15}$$

is fulfilled. This condition allows to calculate the excess temperature that the polariton gas should possess in order to allow the condensation effect to take place. Figure 1 shows the plot of $T_X - T$ versus $T$ for realistic CdTe microcavity parameters, as in [25] (2λ-cavity with a Rabi splitting 26 meV, at positive detuning of 7 meV; $W_0$ corresponds to a decay time of 2.8 ps, observed in ref.[29]) One can see that the temperature difference decays versus the lattice temperature, in agreement with the experimental observations [25]. If the condition (15) is not fulfilled, the gain condition will be achieved for a state having a finite nonzero wave vector. This is an illustration of the bottleneck effect and of the so-called kinetic regime [25]. Accumulation of particles in one state is governed by the relaxation kinetics of the particles and no more by the thermodynamic principles. In that case, the hypothesis of thermal equilibrium that we make breaks down and our approach is no more valid. The fact that stimulation can take place in one or possibly several excited states simply results from the peculiar wave vector dependence of the life time in this simplified model. The complex wave vector dependence of scattering rates is also playing a strong role in realistic situations.

In the following we will assume that $\beta > 0$ and that the eq. (15) is verified, which in the language used in previous works [25] means that the system is in the thermodynamic regime. In such a case, the equation 2 allows as a solution the homogeneous wave function given by the formula 8 with a chemical potential given by:

$$\mu = \alpha \left( |\psi_0|^2 + n_R \right) \tag{16}$$

Contrary to some other models [9], plane wave solutions with finite wave vectors are not the stationary solutions of the system of equations in that case. We now apply the standard linearization procedure (Bogoliubov approximation) in order to find the dispersion of the elementary excitations of the condensate. We therefore write:

$$\psi(\vec{r},t) = \psi_0 \exp(-i\mu t)\left(1 + A_{\vec{k}} \exp\left(i(\vec{k}\vec{r} - \omega_k t)\right) + B^*_{\vec{k}} \exp\left(-i(\vec{k}\vec{r} - \omega_k^* t)\right)\right) \tag{17}$$

and we neglect interactions between the small excitations which mean that we keep only first order terms in $A_k$ and $B_k$. Here we reach a central aspect of this work. In [9] the wave (17) was inducing a spatial modulation of the reservoir density. Since stimulated scattering was written in real space and not in reciprocal space, this spatial modulation of the gain provoked the appearance of a diffusive mode close to $k = 0$. In the present model stimulated scattering is written in reciprocal space and results from a summation over real space. If phonon scattering only is considered, the perturbation (17) is not coupled to the normal phase in the first order and the scattering rates (9) and (10) remain unchanged. If polariton-polariton interaction is considered, the spatial modulation (17) can in principle induce a spatial modulation of the reservoir density because of the scattering rates (5). However, the perturbation (17) cannot induce a change of the distribution of particles in reciprocal space. The condensate distribution function is unchanged ($|\psi(0)|^2 = |\psi_0|^2$ and $|\psi(k)|^2 \sim |A_k|^2, |B_k|^2 \approx 0$) and therefore the distribution function of the normal phase $n(k) = \sum_r n(k,r)$ is unchanged. As a result, the modulation of the scattering rates (9) and (10) induced by the perturbation (17) can be neglected. Finally we obtain the dispersion of the elementary excitations of the condensate:

$$\hbar \omega_k = i(\Delta W_k - \gamma_k) \pm \sqrt{\left(\frac{\hbar^2 k^2}{2m}\right)^2 + 2\alpha^2 |\psi_0|^4 \left(\frac{\hbar^2 k^2}{2m}\right)} \qquad (18)$$

Close of $k=0$, (18) reads:

$$\hbar \omega_k \approx \pm \hbar c_s k - i\beta \frac{\hbar^2 k^2}{2m} \qquad (19)$$

This is a Bogoliubov-like dispersion, which, according to the Landau criterion of superfluidity, demonstrates that these excitations can propagate in the media without being scattered by structural disorder. The speed of sound $c_s = \alpha |\psi_0|^2$, is given by the condensate density and not by the value of the chemical potential. The second difference with respect to the infinite life time case is that elementary excitations of the condensate decay in time with a rate given by $\beta \frac{\hbar^2 k^2}{2m}$ which quadratically grows with $k$, which means that these modes are decaying in time and that the decay is faster for perturbations propagating with a large wave vector. This decay is due to the imbalance between the stimulated scattering process and the life time of a state when $k$ is non zero. This imbalance leads to a decay in time of a superfluid wave travelling in the media. This decay can be seen as a dissipation process but the origin is not a mechanical friction. The velocity of the particles belonging to a superfluid flow remains constant, but their number is decreasing in time. The system is therefore expected to demonstrate physical phenomena typical for a superfluid. In a real system, having a finite size and finite fluctuations, the condensate itself is possessing a finite coherence time, calculated theoretically in several works [17] and measured in the hundreds of ps scale [30]. This coherence time is comparable with the intrinsic decay time we find here. The superfluid fraction is given by the ratio $\frac{|\psi_0|^2}{n_R + |\psi_0|^2}$.

In conclusion, we propose a closed set of equations to describe BEC of weakly interacting in an open system at finite temperature. This model is applied to the specific case of cavity polaritons. It takes properly into account important physical processes neglected in other models extensively used in the literature, such as the thermalisation processes and the wave vector dependence of the life time. We then analyse the specific case of an infinitely large homogeneous system assuming that normal states can be described by a thermal Boltzmann function. The critical condition for the formation of a condensed phase is derived. We show that this condensed phase shows a superfluid response to mechanical excitations with a renormalized sound velocity compared to the infinite life time case. The propagating superfluid waves are also found to decay in time. The results of our model are in sharp contrasts with the one of previous works [8,9] where it was claimed that a Bose condensate in an open system with pumping and decay should not show the Bogoliubov-like spectrum of a superfluid. Our model supports the recent claiming of observations of such Bogoliubov like spectrum under non resonant pumping in a GaAs cavity [10]. We expect that further experimental results will come to support this important conclusion. If the disorder is small enough, there are no fundamental obstacles to observe a Kosterlitz-Thouless phase transition towards a superfluid state in an open system, provided that the relaxation processes are fast enough to allow the formation of a thermal distribution function.

Aknowledgements: The authors wish to thank R. Johne, G. Pavlovic, H. Flayac, I.A. Shelykh, N. A. Gippius, A. V. Kavokin, and J. Bloch. We acknowledge the support of the EU FP7 Spin-Optronics project.

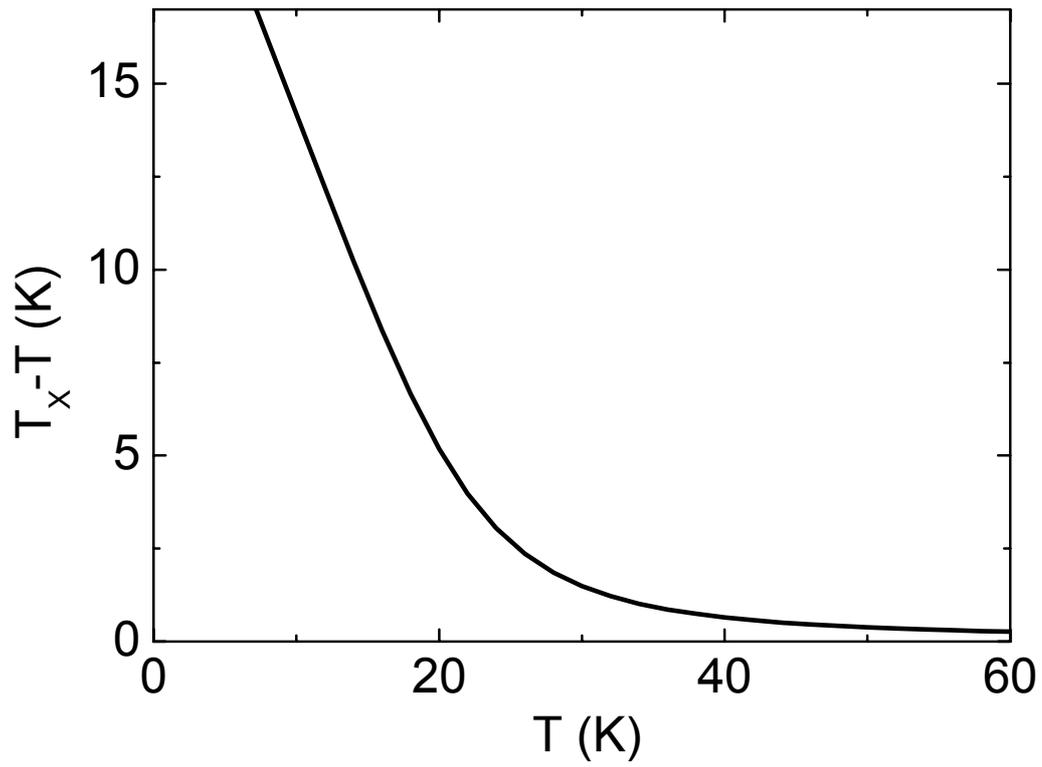

Figure 1. Difference between the temperatures of the polariton gas $T_x$ and of the lattice $T$ versus the lattice temperature $T$, found from the gain condition for the condensate wave function.